\documentclass[preprint,aip,jcp]{revtex4-1}
\usepackage[version=3]{mhchem}
\usepackage{titlecaps}
\usepackage{chemformula} 
\usepackage[T1]{fontenc} 
\usepackage{enumitem}
\usepackage{graphicx}
\usepackage{bm}

\Addlcwords{a an the of on at with for and in into from by without}

\usepackage{xr}
\makeatletter
\newcommand*{\addFileDependency}[1]{
 \typeout{(#1)}
 \@addtofilelist{#1}
 \IfFileExists{#1}{}{\typeout{No file #1.}}
}
\makeatother

\newcommand*{\myexternaldocument}[1]{%
    \externaldocument{#1}%
    \addFileDependency{#1.tex}%
    \addFileDependency{#1.aux}%
}
\myexternaldocument{si}

\begin{document}

\author{Sarbani Patra}
\affiliation{Department of Chemistry, University of Basel,
  Klingelbergstrasse 80, CH-4056 Basel, Switzerland}

\author{Juan Carlos San Vicente Veliz}
\affiliation{Department of Chemistry, University of Basel,
  Klingelbergstrasse 80, CH-4056 Basel, Switzerland}

\author{Debasish Koner}\affiliation{Department of Chemistry,
  University of Basel, Klingelbergstrasse 80, CH-4056 Basel,
  Switzerland}\affiliation{Department of Chemistry, Indian Institute
  of Science Education and Research (IISER) Tirupati, Karakambadi
  Road, Mangalam, Tirupati 517507, Andhra Pradesh, India}

\author{Evan J. Bieske} \affiliation{Department of Chemistry,
  University of Melbourne, Parkville 3010, Australia}

\author{Markus Meuwly}\email{m.meuwly@unibas.ch}
\affiliation{Department of Chemistry, University of Basel,
  Klingelbergstrasse 80, CH-4056 Basel, Switzerland}

\date{\today}

\title[]{Photodissociation dynamics of N$_{3}^{+}$}

\begin{abstract}
The photodissociation dynamics of N$_3^+$ excited from its $^3
\Sigma_{\rm g}^{-}$ ground to the first excited singlet and triplet
states is investigated. Three dimensional potential energy surfaces
for the $^1$A$'$, $^1$A$''$, and $^3$A$'$ electronic states,
correlating with the $^1 \Delta_{\rm g}$ and $^3 \Pi_{\rm u}$ states
in linear geometry, for N$_3^+$ are constructed using high level
electronic structure calculations and represented as reproducing
kernels. The reference {\it ab initio} energies are calculated at the
MRCI+Q/aug-cc-pVTZ level of theory. For following the
photodissociation dynamics in the excited states, rotational and
vibrational distributions $P(v')$ and $P(j')$ for the N$_2$ product
are determined from vertically excited ground state distributions. Due
to the different shapes of the ground state $^3$A$^{''}$ PES and the
excited states, appreciable angular momentum $j' \sim 60$ is generated
in the diatomic fragments. The lifetimes in the excited states extend
to at least 50 ps. Notably, results from sampling initial conditions
from a thermal ensemble and from the Wigner distribution of the ground
state wavefunction are comparable.
\end{abstract}

\maketitle

\section{Introduction}
Many important processes occurring in the earth's atmosphere involve
nitrogen-containing species as it is the most abundant element in the
medium.  Among these are charge transfer processes such as N($^{4}$S)
+ N$_2^+$(X$^2 \Sigma_{\rm g}^+$) $\leftrightarrow$ N$^+$($^3$P) +
N$_2$(X$^1 \Sigma_{\rm g}^+$) that proceed via the formation of the
N$_3^+$ radical cation. An early study utilized the complete active
space self-consistent field (CASSCF) approach and multireference
configuration interaction method (MRCI) to calculate vertical
excitation energies and specify the collinear dissociation paths of
the electronically excited states of the N$_{3}^+$
ion.\cite{rosmus.n3:1996} Given its importance in a variety of
atmospheric reactions, the photodissociation dynamics of the
$\textrm{N}_{3}^{+}$ ion warrants a detailed investigation.\\

\noindent
Reactions involving N$_3^+$ are relevant in air plasmas at elevated
temperatures. Under such conditions, N$_3^+$ can serve as an N$^+$
donor to species such as NO, O$_2$, SO$_2$, N$_2$O, CO$_2$, or
CO.\cite{calvaresi:2006,viggiano.n3p:2004,dunkin:1971,popovic:2004}
Thus, decomposition of N$_3^+$ into N$^+$ + N$_2$ is of particular
relevance in view of the general reaction scheme N$_3^+$ + M
$\rightarrow$ NM$^+$ + N$_2$ with subsequent decomposition of
NM$^+$. Species of particular importance that are formed through N$^+$
transfer reactions are NCO$^+$ (N$_3^+$ + CO $\rightarrow$ NCO$^+$ +
N$_2$) and NCO$_2^+$ (N$_3^+$ + CO$_2$ $\rightarrow$ NCO$_2^+$ +
N$_2$).\cite{calvaresi:2006} Furthermore, N$_3^+$ together with
N$_4^+$ have been suggested to be the most abundant
nitrogen-containing ions in the lower atmosphere of Titan where the
N$_3^+$ ion drives several reactions involving CH$_4$, C$_2$H$_2$ or
C$_2$H$_4$ through N$^+$ atom transfer.\cite{mcewan:2000}\\

\noindent
Photodissociation reactions are induced by the absorption of one or
more photons by a chemical species which can then disintegrate to form
products. In a typical photolytic reaction, a reactant ABC with
internal energy $E_{\rm int}$ absorbs a photon with energy $h\nu$ to
form a transient activated complex [ABC]* which can undergo several
types of primary photochemical processes followed by secondary
processes. One possible outcome is the subsequent dissociation of
[ABC]* into A+[BC]* where the product BC molecule is formed in a
variety of excited quantum states whose population depends on the
available energy and the dynamical details of the dissociation
process\\
\begin{equation}
{\rm ABC} \overset{h\nu} {\rightarrow} {\rm [ABC]}^{*} \rightarrow
{\rm A+[BC]*}
\end{equation}
\\

\noindent
A dynamics study for a chemical reaction usually begins with the
generation of an accurate potential energy surface (PES) for the
species. For the system of interest, 1- and 2-dimensional sections
through the 3-dimensional PESs for the ground and electronically
excited states have been previously determined at different levels of
theory.\cite{rosmus.n3:1994,rosmus.n3:1996,tosi:2004} In addition,
one- and 3-dimensional PESs for cyclic-N$_3^+$ were
determined,\cite{yarkony:2007,krylov:2006,krylov:2006.2} and the
vibrational levels up to $\sim 2$ eV were
computed.\cite{krylov:2006,krylov:2006.2} Finally, {\it ab initio} MD
simulations for the vibrational spectroscopy of linear and cyclic
N$_3^+$ have been carried out.\cite{jolibois:2009} These studies were
concerned with the electronic spectroscopy and the lower vibrational
states on the ground state PES. However, a full-dimensional PES for
the electronic ground state has only become available
recently.\cite{MM.n3p:2020}\\

\noindent
The construction of high-dimensional PESs is a challenging problem for
which machine learning-based approaches like neural networks or
kernel-based methods have found wide applicability in recent
times.\cite{MM.rkhs:2017,manzhos:2021,unke:2021,mm.rev:2021} One such
method, rooted in the theory of reproducing kernel Hilbert spaces
(RKHSs),\cite{rabitz:1996} is suitable for obtaining reliable
PESs. For this, a training set of data is generated using electronic
structure calculations which is then used to "train" the algorithm to
produce a continuous surface by interpolating smoothly between the
data points. Here this 'training set' constitutes the total electronic
energy evaluated at various configurations using high-level electronic
structure calculations. This method was utilized in previous work for
the construction of an accurate PES for the ground state of
$\textrm{N}_{3}^{+}$ at the multi reference configuration interaction
(MRCI) level of theory.\cite{MM.n3p:2020} A similar methodology has
been followed for the construction of high quality PESs for other
triatomic species such as the [CNO] system,\cite{MM.cno:2018}
N$_{3}^{-}$,\cite{salehi.n3m:2019} and NO$_2$.\cite{MM.no2:2020} In
the current work the focus is on those excited states of the N$_3^+$
cation which are expected to be accessible during
photoexcitation. Photodissociation reactions of the N$_3^+$ ion
involve transitions between the $^3 \Sigma_{\rm g}^-$ ($^3$A$''$ in
bent geometry) ground state and energetically accessible excited
states. The $^1$A$'$, $^1$A$''$ and $^3$A$'$ excited states of
N$_{3}^{+}$ are expected to play a role in dynamical processes owing
to their energetic proximity to the ground state ($^3 \Sigma_{\rm
  g}^-$).\\

\noindent
The present work is structured as follows. First, the methods are
presented. Next, the quality of the PESs is discussed and the results
from the dynamics simulations for photoexcitation to the $^3$A$^{'}$
and the two singlet states are reported. Next, simulations starting
from sampling the Wigner distribution of the ground state wavefunction
are compared with the more conventional thermal initial
conditions. Finally, the results are discussed and conclusions are
drawn.\\

\section{Methods}
First, the methods for generating and representing the potential
energy surfaces are summarized. Following this, the QCT and quantum
simulations for studying N$_3^+$$\rightarrow$N$_2$+N$^+$
photodissociation are described.\\

\subsection{The N$_3^+$ Potential Energy Surfaces}
The 3-dimensional PESs for the $^1$A$'$, $^1$A$''$, and $^3$A$'$
states of N$_3^+$ were computed at the
MRCISD\cite{wer88:5803,kno88:514}/aug-cc-pVTZ\cite{dun89:1007} level
of theory with the Davidson quadruples correction\cite{davidson:1974}
(MRCISD+Q) based on a
CASSCF\cite{wen85:5053,kno85:259,wer80:2342,werner:2019} reference
wave function. Also, the 2d-PES at $r_{\rm NN} = 2.25$ a$_0$ (see
below) for the $2 ^3$A$^{''}$ state was determined. All electronic
structure calculations were carried out using the Molpro 2019.1
program.\cite{molpro,MOLPRO_brief} The active space for CASSCF
included the full valence space. State-averaged (SA) calculations were
carried out using 8 states in total, including the two lowest states
of the $^3$A$''$ (ground state), $^1$A$'$, $^1$A$''$, and $^3$A$'$
symmetries, respectively.\\

\noindent
For the electronic structure calculations, the grid was defined in
Jacobi coordinates $(R, r, \theta)$ whereby $r$ is the separation
between nitrogen N1 and N2, $R$ is the distance between N3 and the
center of mass of N1 and N2, and $\theta$ is the angle between
$\vec{r}$ and $\vec{R}$; see Figure \ref{fig:n3ppes}. The grid
includes distances $r \in [1.55,4.00]$ a$_0$, $R \in [1.5,10.0]$
a$_0$, and $\theta \in [0, 180^\circ]$ from a $7$-point Legendre
quadrature. The products of the photodissociation reaction of N$_3^+$
are N$_2$ and N$^+$. Thus, the energy of the system for $E(R
\rightarrow \infty,r)$ is the sum of the energies of the dissociation
products. In the following, the "zero" of energy is set to this value
according to $V(R,r,\theta)=E(R,r,\theta)- E(R\rightarrow
\infty,r)$. \\

\noindent
For the dynamics simulations the energies on the grid are most
conveniently represented in a way that allows evaluation of energies
and analytical gradients for arbitrary conformations. Here, a
reproducing kernel Hilbert space (RKHS) representation is
employed.\cite{rabitz:1996,MM.rkhs:2017} According to the representer
theorem\cite{scholkopf2001generalized}, a function $f(\mathbf{x})$ for
which values $f(\mathbf{x_i})$ are given for arguments $\mathbf{x}$,
can always be approximated as a linear combination
\begin{equation}
f(\mathbf{x}) \approx \widetilde{f}(\mathbf{x}) = \sum_{i = 1}^{N}
\alpha_i K(\mathbf{x},\mathbf{x}_i).
\label{eq:kernel_regression}
\end{equation}
Here, $K(\mathbf{x},\mathbf{x'})$ is a kernel function and $\alpha_i$
are coefficients to be determined from matrix inversion. If the inner
product $\langle \phi(\mathbf{x}),\phi(\mathbf{x'})\rangle$ can be
written as $K(\mathbf{x},\mathbf{x'})$, the function is called a
reproducing kernel of a Hilbert space
$\mathcal{H}$.\cite{berlinet2011reproducing}\\

\noindent
Here, reciprocal power decay kernel polynomials are used for the
radial coordinates. For the $R-$coordinate kernel functions
($k^{[n,m]}$) with smoothness $n=2$ and asymptotic decay $m=4$
\begin{equation}
\label{k24}
 k^{[2,4]}(x,x') = \frac{2}{15}\frac{1}{x^5_{>}} -
 \frac{2}{21}\frac{x_<}{x^6_>},
\end{equation}
are employed, while $n=2$ and $m=6$ is used for the $r$ dimension
\begin{equation}
 k^{[2,6]}(x,x') = \frac{1}{14}\frac{1}{x^7_{>}} -
 \frac{1}{18}\frac{x_<}{x^8_>},
\end{equation}
In both expressions, $x_>$ and $x_<$ are the larger and smaller values
of $x$ and $x'$, respectively. Such a kernel smoothly decays to zero
giving the correct long-range behavior for atom-diatom type
interactions. For the angle $\theta$ a Taylor spline kernel is used:
\begin{equation}
 k^{[2]}(z,z') = 1 + z_<z_> + 2z^2_<z_> - \frac{2}{3}z^3_<,
\end{equation}
Here, $z_>$ and $z_<$ are the larger and smaller values of $z$ and
$z'$, respectively, and $z$ is defined as
\begin{equation}
z = \frac{1 - {\rm cos} \theta}{2},
\end{equation}
with $z \in [0,1]$. Combination of the three 1-dimensional kernels
leads to
\begin{equation}
\label{3dk}
 K({\bf{x}}, {\bf{x}}') = k^{[2,4]}(R,R')k^{[2,6]}(r,r')k^{[2]}(z,z'),
\end{equation}
where, ${\bf{x}}, {\bf{x}}'$ are $(R, r, z)$ and $(R', r', z')$,
respectively. The coefficients $\alpha_i$ and the RKHS representation
of the PES are evaluated by using a computationally efficient
toolkit.\cite{MM.rkhs:2017}\\

\noindent
The global reactive PES for each excited electronic state is
constructed by smoothly connecting the three PESs for the three
symmetry-equivalent reaction channels using a switching function,
\begin{equation}
    V(\mathbf{r})= \sum_{j=1}^{3} \omega_{j} (r_{j}) V_{j}(\mathbf{r})
\end{equation}
where the switching function $\omega_{j}$ has an exponential form,
\begin{equation}
    \omega_{i}(r)=\frac{e^{-(r_{i}/\rho_j)^2}}{\sum_{j=1}^{3}e^{-(r_{j}/\rho_j)^2}}
\end{equation}

\noindent
Such a mixing using normalized weights is akin to that used in multi
surface reactive MD.\cite{mm.msarmd:2014,mm.msvalbond:2018} The mixing
parameters $\rho_j$ for each channel are obtained using least squares
fitting. For the $^3$A$^{''}$ ground state of $\textrm{N}_{3}^{+}$,
the switching parameters are $\rho =(0.65,0.65,0.65)$
a$_{0}$.\cite{MM.n3p:2020} The root mean squared error between the two
sets of data is $\sim 0.8$ kcal/mol (0.034 eV) over the entire range
of reference energies considered. Similarly, reactive PESs were
constructed for the $^1$A$^{'}$, $^1$A$^{''}$ and $^3$A$^{'}$ excited
states. For the $^1$A$^{'}$ state, the switching parameters are $\rho
= (1.0,1.0,1.0)$ a$_0$ with an RMSE of $\sim 0.6$ kcal/mol (0.026
eV). For the $^1$A$^{''}$ state, the switching parameter was $\rho =
(1.03,1.03,1.03)$ a$_0$ with an RMSE of $\sim 0.6$ kcal/mol (0.026 eV)
and for the $^3$A$^{'}$ state, a switching parameter of $\rho =
(0.75,0.75,0.75)$ a$_0$ yielded an RMSE of $\sim 0.06$ kcal/mol (0.003
eV) for the global reactive PES.\\

\subsection{Quasi-Classical Trajectory Simulations}
The QCT simulations used in the present work have been extensively
described in the literature.\cite{tru79,hen11,kon16:4731,MM.cno:2018}
Here, Hamilton's equations of motion are solved using a fourth-order
Runge-Kutta numerical method. The time step was $\Delta t = 0.05$ fs
which guarantees conservation of the total energy and angular
momentum.\\

\noindent
For the photodissociation simulations, structures in the vicinity of
the $^3$A$''$ ground state PES and velocities were generated by
drawing from a Maxwell-Boltzmann distribution at temperatures between
500 K and 3000 K. Each of the 500000 initial conditions was propagated
for 50 ps on the ground state PES and the final positions and
velocities were saved. For a view of the ensemble of structures; see
Figure \ref{sifig:incond_proj_3ap}. Following this, the entire
population is projected vertically to the excited state
PESs. Trajectories on the excited states are run until dissociation
into products occurs or for a maximum of 50 ps. Configurations
initially located around the ground state minima land in the vicinity
of a potential well when projected onto the $^{3}\textrm{A}^{'}$
PES. Subsequent trajectories are initially confined in the region
around the minima before dissociating into products. Examples for
photodissociating trajectories from initial velocities generated at
1000 K on the $^{3}\textrm{A}^{'}$ PES are shown in Figure
\ref{sifig:pdtraj_3ap}. \\

\noindent
The product ro-vibrational states are determined following
semiclassical quantization. Since the ro-vibrational states of the
product diatom are continuous numbers, the states need to be assigned
to integer quantum numbers for which a Gaussian binning (GB) scheme
was used. For this, Gaussian weights centered around the integer
values with a full width at half maximum of 0.1 were
used.\cite{bon97:183,bon04:106,kon16:4731} It is noted that using
histogram binning (HB) was found to give comparable results for a
similar system.\cite{MM.cno:2018}\\

\subsection{Bound Vibrational States for Electronically Excited States of N$_3^+$ }
The vibrational energy levels supported by the singlet excited state
PESs are computed using the DVR3D suite of codes.\cite{dvr3d:2004} For
this, the nuclear time-independent Schr\"odinger equation is solved
over a discrete grid in Jacobi coordinates $(R, r, \theta)$. In this
method, the three internal coordinates are treated in a discrete
variable representation (DVR). The angular coordinate is represented
as a 56-point DVR based on Gauss-Legendre quadrature and the radial
coordinates utilise a DVR based on Gauss-Laguerre quadratures with 72
points along $R$ and 48 points along $r$. The angular basis functions
are Legendre polynomials and the radial basis functions are Laguerre
polynomials. For the $^3$A$^{'}$ PES, the states predissociate due to
the double-well structure of the surface; see Figure
\ref{fig:n3p_1d}.\\

\noindent
The optimized Morse parameters for the grid in $r$ for the $^1$A$'$
state are $r_e = 2.50$ a$_0$, $D_e = 0.32$ $E_{\rm h}$ and $\omega_e =
0.006$ $E_{\rm h}$, and $R_e = 3.4$ a$_0$, $D_e = 0.15$ $E_{\rm h}$
and $\omega_e = 0.0015$ $E_{\rm h}$ for $R$. With these parameters the
$r$ grid for $^1$A$'$ extends from 1.49 to 3.46 a$_0$ while the $R$
grid ranges from 1.20 to 5.51 a$_0$. The corresponding Morse
parameters for the $^1$A$''$ state are $r_e = 2.45$ a$_0$, $D_e =
0.32$ $E_{\rm h}$ and $\omega_e = 0.0065$ $E_{\rm h}$ along $r$ and
$R_e = 3.5$ a$_0$, $D_e = 0.15$ $E_{\rm h}$ and $\omega_e = 0.0015$
$E_{\rm h}$ for $R$. The $r$ grid is from 1.48 to 3.36 a$_0$ while the
$R$ grid spans 1.30 to 5.61 a$_0$. For determination of the rotational
levels, the body-fixed $z-$axis is oriented along $R$. The
corresponding vibrational wave functions obtained as amplitudes over a
discrete grid in Jacobi coordinates are transformed to symmetric
$(R_{\rm N1N2}+R_{\rm N2N3})/\sqrt{2}$), asymmetric $(R_{\rm
  N1N2}-R_{\rm N2N3})/\sqrt{2}$) and bending coordinates $\angle$
N1N2N3 using a Gaussian kernel-based interpolation method. Here, the
triatomic is denoted as N1--N2--N3. This transformation allows the
approximate assignment of the wavefunctions using quantum numbers
obtained by counting the nodal planes along each coordinate.\\

\section{Results}
\subsection{The Potential Energy Surfaces of the Excited States}
\begin{figure}[htbp]
\begin{center}
\includegraphics[width=1.0\textwidth]{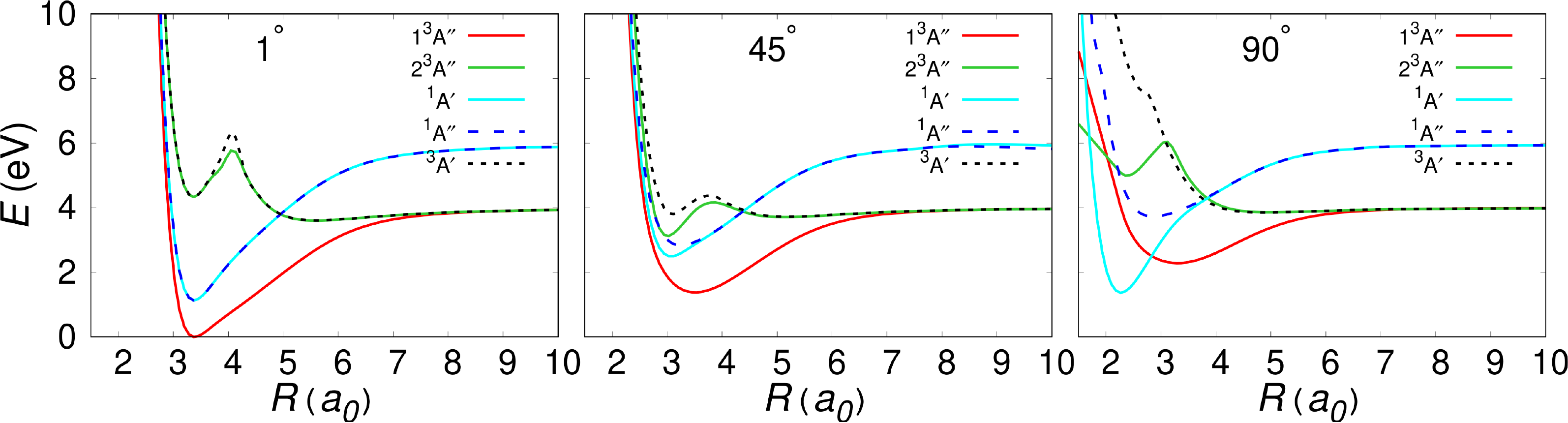}
\caption{One dimensional sections of the three dimensional PESs of the
  ground and excited states of N$_{3}^{+}$. All cuts are taken at
  $r=2.25$ $a_0$ and the values of the angle $\theta$ are indicated at
  the top of each plot. For linear N$_3^+$ ($\theta = 0$) the ground
  state ($^3$A$^{''}$) correlates with the $^3 \Sigma_{\rm g}^{-}$
  state, the two $^1$A states ($^1$A$^{'}$ and $^1$A$^{''}$,
  degenerate in linear geometry) with the $^1 \Sigma_{\rm g}^{+}$
  state, and $^3$A$^{'}$ and $2 ^3$A$^{''}$ correlate with the $^3
  \Pi_{\rm u}$ state.}
  \label{fig:n3p_1d}
\end{center}
\end{figure}

In the following the quality and topology of the excited state PESs
determined in the present work is described. The $1^3$A$^{''}$ ground
state PES, correlating with the $^3 \Sigma_{\rm g}^-$ state in linear
geometry and dissociating to N$^+ (^3 {\rm P})$+N$_2 ({\rm X} ^1
\Sigma_{\rm g}^+ )$, was already discussed in earlier
work.\cite{MM.n3p:2020} Next up in energy are the $^1$A$^{'}$ and
$^1$A$^{''}$ states (see left panel in Figure \ref{fig:n3p_1d}), which
are degenerate for $\theta = 0$, correlate with the $^1 \Delta_{\rm
  g}$ state and dissociate to N$^+ (^1 {\rm D})$+N$_2 ({\rm X} ^1
\Sigma_{\rm g}^+ )$. Upon bending away from the linear structure the
degeneracy is lifted and two states - $^1$A$^{'}$ and $^1$A$^{''}$ -
emerge as is shown in Figure \ref{fig:n3p_1d} middle and right
panels. The overall shapes of the two PESs for the $^1$A states follow
that of the $^3$A$^{''}$ ground state PES.\\

\noindent
At yet higher energy and most relevant for the photodissociation
dynamics considered later, are the $^3$A$^{'}$ and $2^3$A$^{''}$
states which correlate with the $^3 \Pi_{\rm u}$ state in linear
geometry. Both states dissociate to N$^+ (^3 {\rm P})$+N$_2 ({\rm X}
^1 \Sigma_{\rm g}^+ )$, i.e. they have the same asymptote as the
$1^3$A$^{''}$ ($^3 \Sigma_{\rm g}^-$) ground state. The $^3$A$^{'}$
state has a double minimum PES with a first minimum at $R = 3.365$
a$_0$ which is higher in energy than the corresponding N$_2$($^1
\Sigma_{\rm g}^{+}$)+N$^+$($^3$P) asymptote by 0.328 eV and separated
from it by a barrier of 2.351 eV in the linear configuration; see
Figure \ref{fig:n3p_1d}. This double well structure of the PES
disappears upon bending of the N$_3^+$ molecule and leads to a
repulsive PES for a T-shaped structure, see right hand panel in Figure
\ref{fig:n3p_1d}. Thus, $^3 \Pi_{\rm u}$ $\leftarrow$ $^3 \Sigma_{\rm
  g}^-$ excitation of linear N$_3^+$ from the ground to the excited
state is expected to lead to pronounced angular dynamics, also because
the structure with $\theta = 45^\circ$ in the region of the excitation
($R \in [3 \cdots 4]$ a$_0$) is lower in energy than for the linear
conformation, see middle panel Figure \ref{fig:n3p_1d}. For the $2
^3$A$^{''}$ state the 1d-cuts resemble the $^3$A$^{'}$ state for the
linear and $\theta = 45^\circ$ geometries. For the T-shaped
conformation there is a local minimum at $R = 2.384$ a$_0$. For $R >
3$ a$_0$ the $^3$A$^{'}$ and $2 ^3$A$^{''}$ states overlap
again. Because the geometry of the $^3$A$^{''}$ ground state is
linear, photoexcitation to the triplet states is expected to primarily
populate regions around $\theta = 0$.\\

\noindent
The vertical excitation energies for the $^1 \Delta_{\rm
  g}$$\leftarrow$$^3 \Sigma_{\rm g}^-$ and the $^3 \Pi_{\rm
  u}$$\leftarrow$$^3 \Sigma_{\rm g}^-$ transitions have been
determined or estimated from experiments.\cite{dyke:1982} For the $^1
\Delta_{\rm g}$$\leftarrow$$^3 \Sigma_{\rm g}^-$ transition the
excitation energy estimated from photoelectron spectroscopy is 1.13
eV, and for the $^3 \Pi_{\rm u}$$\leftarrow$$^3 \Sigma_{\rm g}^-$
transition a laser spectroscopy study yielded an excitation energy of
4.54 eV.\cite{maier.n3p:1994} From the present calculations and for
the linear configuration with $r = 2.25$ a$_0$, these two excitation
energies are 1.13 eV and 4.35 eV, respectively; see Figure
\ref{fig:n3p_1d}. These results agree favourably with
experiments. Earlier MRCI calculations reported transition energies of
1.30 eV and 4.48 eV and 1.13 eV and 4.28 eV when Davidson corrections
are included.\cite{rosmus.n3:1996}\\

\begin{figure}[htbp]
\includegraphics[width=0.95\textwidth]{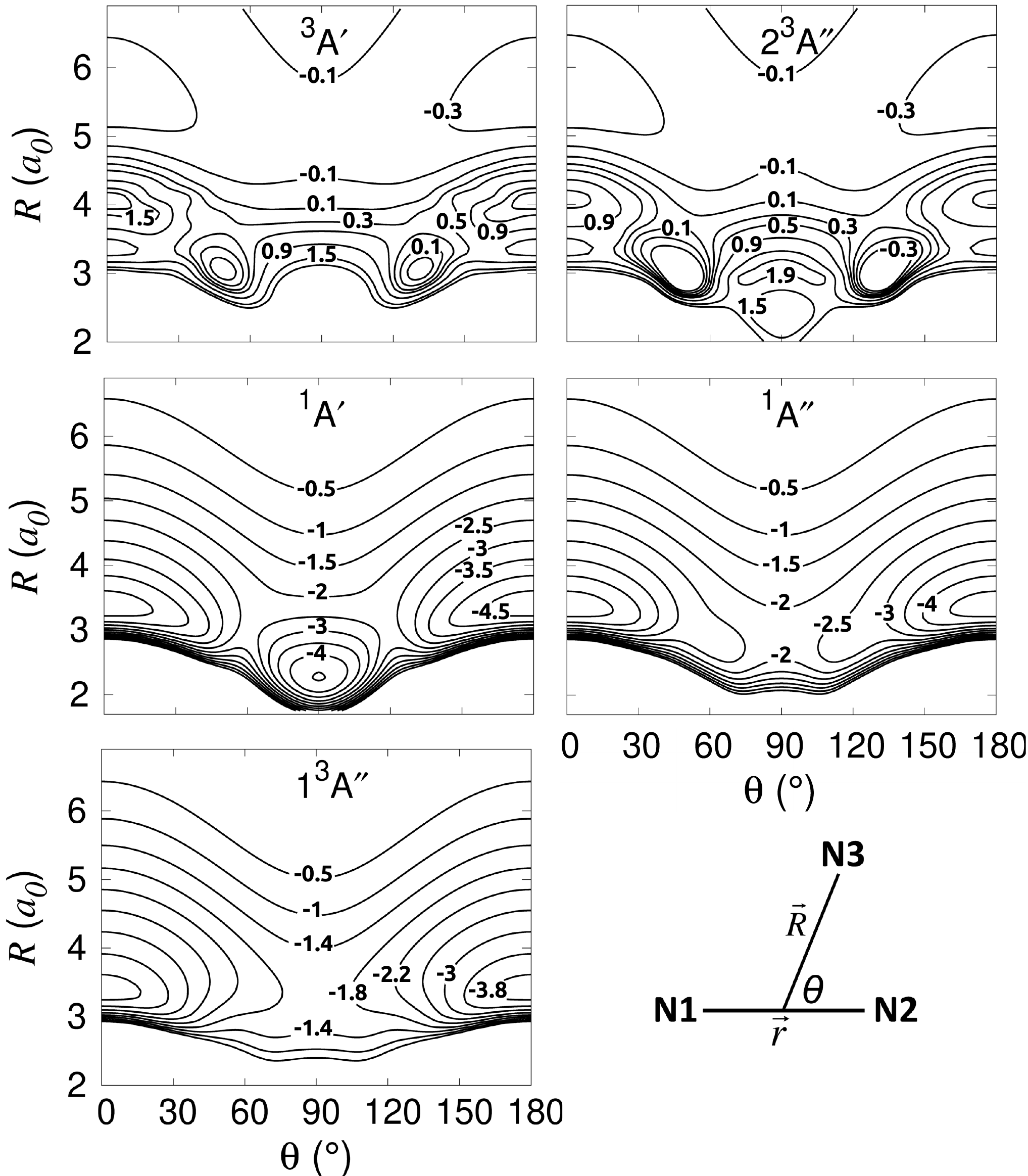}
\caption{Two-dimensional contour plots for the PESs of the $1
  ^3$A$^{''}$ ground state ($^3 \Sigma_{\rm g}^{-}$ for linear
  N$_3^+$, lower left), the two $^1$A states ($^1$A$^{'}$, middle left
  and $^1$A$^{''}$ middle right) corresponding to the $^1 \Delta_{\rm
    g}$ state in the linear structure, and the two $^3$A states
  ($^3$A$^{'}$ middle left and $2 ^3$A$^{''}$ middle right)
  correlating with the $^3 \Pi_{\rm u}$ state. In all panels the
  N1--N2 diatomic separation is fixed at $r=2.25$ a$_0$. The asymptote
  for the triplet states is $E_{\rm N_2}+E_{\rm N^+(^3P)}$ and for the
  singlet states it is $E_{\rm N_2}+E_{\rm N^+(^1D)}$. All energies
  are reported in eV. The bottom right panel shows the Jacobi
  coordinates $(R,r,\theta)$ used for describing the 3-dimensional
  PESs for N$_3^+$. The vector $\vec{r}$ is the N1--N2 separation,
  $\vec{R}$ is the separation between N3 and the center of mass of
  N1--N2, and $\theta$ is the angle between the two distance vectors.}
\label{fig:n3ppes}	
\end{figure}

\noindent
The RKHS representations for the four electronically excited states
are provided in Figure \ref{fig:n3ppes}. The minimum energy structure
of the $^3$A$^{''}$ ground state (Figure \ref{fig:n3ppes} bottom left)
is 3.99 eV below the N$_2$($^1 \Sigma_{\rm g}^{+}$)+N$^+$($^3$P) limit
with a barrier of $2.28$ eV for interconversion between the two
equivalent linear minima. The $^3$A$^{'}$ state (Figure
\ref{fig:n3ppes} upper left) has minima at $\theta = 49.8^\circ
(130.8^\circ)$, 0.493 eV above the N$_2$($^1 \Sigma_{\rm
  g}^{+}$)+N$^+$($^3$P) dissociation limit and local minima at $\theta
= 0^\circ (180^\circ)$, 0.328 eV above the same asymptote. It also
contains local maxima at $\theta = 0^\circ (180^\circ)$ with height
2.351 eV above the same asymptote. For the $2 ^3$A$^{''}$ state, a
2-dimensional PES at fixed N1-N2 separation $r_{\rm NN} = 2.25$ a$_0$
was determined on the same $(R,\theta)$ grid as for all other
states. The RKHS representation is illustrated in the upper right
panel of Figure \ref{fig:n3ppes}. For the linear geometry $\theta =
0^\circ$ the $^3$A$^{'}$ and $2 ^3$A$^{''}$ PESs are degenerate and
associated with the $^3 \Pi_{\rm u}$ state (see Figure
\ref{fig:n3p_1d}), with a slight difference around $R \sim 4$
a$_0$. For nonlinear geometries the two states split as was already
found in earlier electronic structure calculations.\cite{tosi:2004}
Overall, the topography of the $^3$A$^{'}$ and $2 ^3$A$^{''}$ states
which both derive from the $^3 \Pi_{\rm u}$ state of linear N$_3^+$
are similar except for a local minimum in the T-shaped geometry for
the $2 ^3$A$^{''}$ state. This minimum is separated by a barrier of
$\sim 1.6$ eV from the minimum at $\theta \sim 45^\circ$. As the
$^3$A$^{'}$ and $2 ^3$A$^{''}$ PESs are similar for $\theta \lesssim
45^\circ$ and photoexcitation from the ground state primarily
populates this region of the PES, the photodissociation dynamics on
the $^3$A$^{'}$ and $2 ^3$A$^{''}$ surfaces are expected to be
comparable. Sampling of the local minimum around $\theta \sim
90^\circ$ on the $2 ^3$A$^{''}$ state following photoexcitation is
unlikely as this local, T-shaped minimum is separated by a barrier
exceeding 1 eV from the region with $\theta \leq 45^\circ$.\\

\noindent
The $^1$A$^{'}$ and $^1$A$^{''}$ states derive from the $^1
\Delta_{\rm g}$ state of linear, centrosymmetric N$_3^+$. The
$^1$A$^{'}$ state (Figure \ref{fig:n3ppes} middle left) has two
minima, a local one at $\theta = 0^\circ (180^\circ)$, 4.824 eV below
the N$_2$($^1 \Sigma_{\rm g}^{+}$)+N$^+$($^1$D) asymptote and the
global one at $\theta = 90^\circ$ which is 5.090 eV below the same
asymptote. The barrier between the two minima is 2.425 eV above the
respective asymptote. The $^1$A$^{''}$ state (Figure \ref{fig:n3ppes}
middle right) features a minimum at linear positions 4.824 eV below
the N$_2$($^1 \Sigma_{\rm g}^{+}$)+N$^+$($^1$D) asymptote. The barrier
between the two symmetrical minima is 2.229 eV above the respective
asymptote. The locations and energies of all the critical points are
summarized in Table \ref{tab:criticalpoints}.\\

\begin{table}[htbp]
\centering
\caption{Minima and transition states for the 3-dimensional ground and
  excited PESs of N$_{3}^{+}$ considered in the present work. The PESs
  are represented in Jacobi coordinates. All interatomic distances are
  in Bohr ($a_0$) and angles are in degree. Energies are reported in
  eV with respect to the
  $\textrm{N}+\textrm{N}+\textrm{N}^+(^3\textrm{P})$ asymptote for the
  triplet states and
  $\textrm{N}+\textrm{N}+\textrm{N}^+(^1\textrm{D})$ asymptote for the
  singlet states. The electronic states are labelled for both, ${\rm
    D}_{\infty \rm h}$ and ${\rm C_s}$ symmetries. Results for the
  $2^3$A$''$ state are not reported because only a 2-dimensional PES
  was determined.}

\begin{tabular}{|l||l|r|r|r|r|}
\hline
                      &        &    $R$   &    $r$     &  $\theta$   &     $E$ \\
\hline\hline
   &  MIN1  &  3.38  &  2.25    &    0.0      & -13.05      \\ 
$^3\Sigma_{\rm g}^{-}$/$^3$A$''$   &  MIN2  &  3.38  &  2.25    &  180.0      & -13.05      \\
            &  TS1   &  3.37  & 2.14     &  90.0       & -10.90     \\
\hline

      &  MIN1  & 2.18   &  2.51    &  90.0        &  -14.10    \\ 
    &  MIN2  & 3.38   & 2.25     &  0.0         &  -13.83     \\
$^1 \Delta_{\rm g}$/$^1$A$'$    &  MIN3  & 3.38   &  2.25    &  180.0       &  -13.83    \\
    &  TS1   & 2.84   & 2.34     &  61.0        &  -11.35    \\
   &  TS2   & 2.84   & 2.34     &  119.0       &  -11.35    \\

\hline 

    &  MIN1  & 3.38   & 2.25     &    0.0       &   -13.83       \\
$^1 \Delta_{\rm g}$/$^1$A$''$    &  MIN2  & 3.38   &  2.25    &  180.0       &   -13.83   \\
    &  TS1   & 2.97   & 2.18     &  90.0       &   -11.27      \\
 
\hline

       &  MIN1  &  3.01  &  2.41  &   49.2   & --9.50  \\
         &  MIN2  &  3.01  &  2.41  &   130.8  & --9.50  \\
      &  MIN3  &  3.36  &  2.24  &    0.0   & --8.69  \\
$^3 \Pi_{\rm u}$/$^3$A$'$      &  MIN4  &  3.36  &  2.24  &   180.0  & --8.69  \\
       &  TS1   &  3.36  &  2.32  &    21.0  & --8.51     \\
       &  TS2   &  3.36  &  2.32  &   159.0  & --8.51     \\
       &  TS3   &  5.06  &  2.09  &    90.0  & --9.56     \\
\hline 

\end{tabular}
\label{tab:criticalpoints}
\end{table}

\begin{figure}[htbp]
\begin{center}
\includegraphics[width=0.93\textwidth]{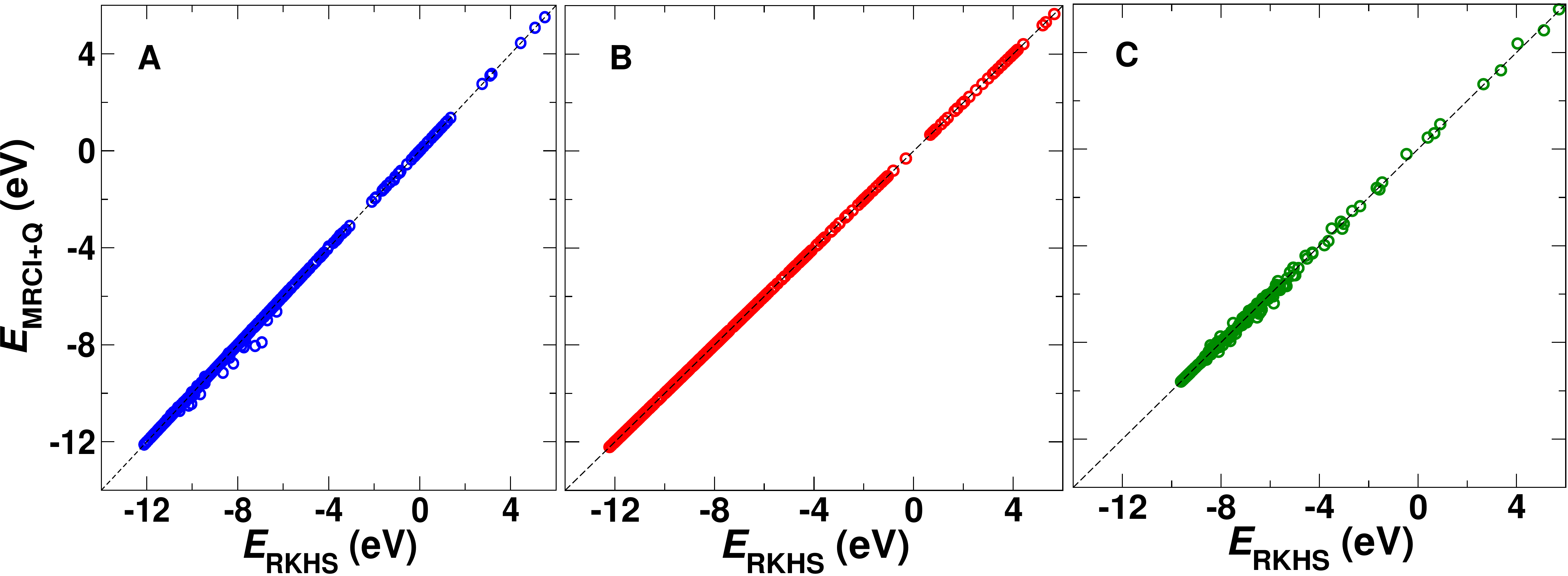}
\caption{Correlation between \textit{ab initio} and RKHS energies for
  the $^1$A$^{'}$ (panel A), $^1$A$^{''}$ (panel B), and $^3$A$^{'}$
  (panel C) excited states of N$_{3}^{+}$. The respective correlation
  coefficients for the three PES are $R^2=0.99969$ for $^1$A$^{'}$,
  $R^2=0.99998$ for $^1$A$^{''}$ and $R^2=0.99978$ for the $^3$A$^{'}$
  state.}
\label{fig:validation}
\end{center}
\end{figure}

\noindent
As a verification for the quality of the RKHS representation,
electronic energies at additional grid points, which are not part of
the training set, are evaluated from {\it ab initio} calculations and
evaluation of the RKHS. The correlation between the reference MRCI+Q
energies and their representation as a RKHS is given in Figure
\ref{fig:validation}. For the $^1$A$^{'}$ state, energies at 400
additional grid points are calculated and a correlation coefficient of
0.99969 is obtained, demonstrating the high accuracy of the
RKHS-represented PESs. Similarly, validation sets of 530 and 315 grid
points for the $^1$A$^{''}$ and $^3$A$^{'}$ states yield correlation
coefficients 0.99998 and 0.99978, respectively. \\

\noindent
The crossing seams between the two singlet electronic states, which
are degenerate in the linear configuration, are reported in Figure
\ref{sifig:n3_crossing}. The geometries at which the two states cross
were stored whenever the energy difference between the $^1$A$^{'}$ and
$^1$A$^{''}$ states was smaller than $10^{-5}$ eV for a given
geometry. Crossing seams are shown for three different values of the
angle $\theta$. Thus, for the excited vibrational states, there is a
possibility of nonadiabatic transitions between the $^1$A$^{'}$ and
$^1$A$^{''}$ states.\\

\subsection{$^3 \Pi_{\rm u}$$\leftarrow$$^3 \Sigma_{\rm g}^{-}$ Photodissociation Dynamics}
\begin{figure}[htbp]
\begin{center}
  \includegraphics[width=1.0\textwidth]{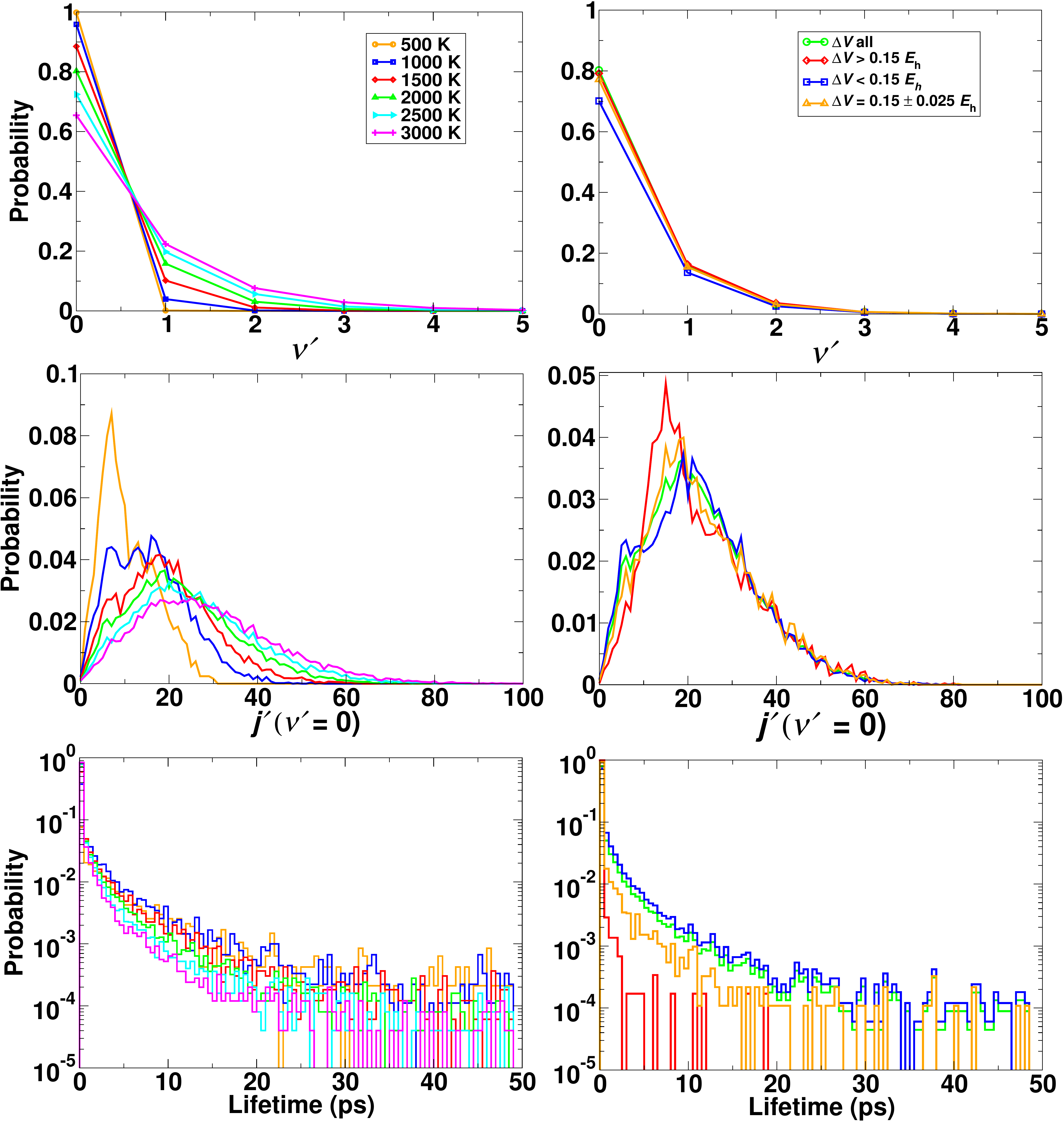}
  	\caption{Photodissociation dynamics on the $^3$A$'$ PES. Left
          column: Product vibrational (top) and rotational (middle)
          state distributions and lifetime distributions (bottom) from
          simulations started with velocities from Maxwell-Boltzmann
          distributions at different temperatures on the ground state
          PES and vertical excitation to the excited state. Right
          column: Product vibrational and rotational state
          distributions and lifetime distributions from simulations
          started with velocities from Maxwell-Boltzmann distributions
          at 2000 K on the ground state PES. The analysis is carried
          out for different criteria for $\Delta V=V_{\rm es}-V_{\rm
            gs}$ as specified in the figure legend and in the text.
          For comparison, the product state distributions when all
          initial conformations undergo photoexcitation are shown in
          blue. In all cases, the resulting trajectories are
          propagated for a maximum time of 50 ps or until
          dissociation, whichever is earlier.}
\label{fig:final3ap}
\end{center}
\end{figure}

The photodissociation dynamics on the excited $^3$A$'$ state is
studied by evaluating the vibrational and rotational state
distributions of the products. The product final state distributions
from initial conditions generated at the different temperatures are
shown in Figure \ref{fig:final3ap}. Vertical photoexcitation of a
thermalized ensemble of molecules on the ground electronic state
results in their promotion to the lowest energy vibrational level of
the excited state. As temperature increases higher $v'$ values become
populated gradually with $v'_{\rm max} = 4$. In Figure
\ref{fig:final3ap} the final rotational distributions at all
temperatures are also shown. Only the results for the most
populated $v' = 0$ state are presented. The excited state population
is distributed over a wide range of rotational states with the peak of
the distribution shifting to higher values as temperature
increases. Examples for photodissociating trajectories at 1000 K on
the $^{3}\textrm{A}^{'}$ PES are shown in Figure
\ref{sifig:pdtraj_3ap}. Configurations initially located around the
ground state minima land in the vicinity of a potential well when
projected onto the $^{3}\textrm{A}^{'}$ PES. Several of these
trajectories are confined in the region around the minima for a few
vibrational periods before dissociating into products. Several
trajectories also arrive at the potential barrier near $\theta \sim
0^\circ$ and undergo dissociation immediately. Note that only
trajectories resulting from initial conditions with $\theta = 0^\circ$
to $90^\circ$ are shown in the figure. Trajectories initialized from
$\theta = 90^\circ$ to $180^\circ$ follow similar dynamical paths on
account of the symmetry of the PES. The distribution of lifetimes
$P(\tau)$ of the $\textrm{N}_{3}^{+}$ complex in Figure
\ref{fig:final3ap} indicates that a large fraction of the trajectories
leads to photodissociation within 5 ps after excitation. However,
there is also a number of trajectories with considerably longer
lifetimes. Figure \ref{sifig:long_lifetime_3Ap} shows one such
trajectory at 1000 K with a lifetime of 17 ps. \\

\noindent
Up to this point the entire ground state population was projected onto
the excited state PES and the dynamics was followed. In other words,
it was assumed that the resonance condition, $h\nu=E_{\rm
  photon}=E_{\rm es}-E_{\rm gs}=\Delta V$, is always fulfilled. Here,
$h\nu$ is the energy of the incoming photon and $E_{\rm gs}$ and
$E_{\rm es}$ are the energies of the molecule in the ground and
excited states, respectively. However, experimentally typically only a
fraction of the population is promoted from the lower to the upper
state. Such processes constitute a subset of the trajectories
discussed so far and are discussed next. An examination of the PESs
corresponding to $\theta \approx 0^\circ$ in Figure \ref{fig:n3p_1d}
reveals that the difference in energies between the respective minima
on the $^3$A$^{'}$ and $^3$A$^{''}$ states is $ \Delta V \approx 4$ eV
indicating that with initial conditions in the vicinity of the ground
state minimum, molecules would require $\sim 4$ eV energy for the
transition to the excited state. Hence, the following different cases
are considered : (1) $ \Delta V > 0.15 E_{\rm h}$ ($\approx 4$ eV),
(2) $\Delta V < 0.15$ $E_{\rm h}$, and (3) $ \Delta V = 0.15 \pm
0.025$ $E_{\rm h}$. The results for the ensuing dynamics from
trajectories sampled from initial conditions at 2000 K are shown in
Figure \ref{fig:final3ap} (right hand column). The respective
distributions when all trajectories are photoexcited are also shown on
the same graphs for comparison. The energy difference for Case (1)
corresponds to a photon wavelength $\lambda<310$ nm. \\

\noindent
If only the low-energy part of the distribution is promoted to the
excited state ($\Delta V < 0.15$ $E_{\rm h}$) the population of the
vibrationally excited state in the product state is slightly smaller
than for the other three cases. Conversely, excitation with $\Delta V
> 0.15$ $E_{\rm h}$ leads to a maximum value $j_{\rm max}'$ which is
somewhat lower than for the remaining cases. The most pronounced
differences arise for the lifetimes on the excited state, which depend
on what fraction of the ground state distribution is excited. For high
energy excitation ($\Delta V > 0.15$ $E_{\rm h}$) lifetimes are
strongly clustered on the picosecond time scales with a maximum
lifetime of 20 ps. For near-resonant excitation ($\Delta V = 0.15 \pm
0.025$ $E_{\rm h}$) lifetimes on the several 10 ps time scale are more
probable, extending out to 50 ps. Excitation of the low-energy part of
the ensemble (green) leads to a higher probability for longer
lifetimes but the shape of the distribution is similar to that for
exciting the entire ground state population (blue).\\

\subsection{$^1 \Delta_{\rm g}$$\leftarrow$$^3 \Sigma_{\rm g}^{-}$
  Photodissociation Dynamics}
Formally, the $^3$A$'$ $\rightarrow$ $^1$A$'$ and $^3$A$'$
$\rightarrow$ $^1$A$''$ transitions involve a change of multiplicity
and are forbidden, and therefore expected to be weak. Nevertheless, it
is of interest to consider how the final state distributions depend on
the different topographies of the underlying PESs by comparing final
state distributions from transitions to the $^3$A$''$, $^1$A$'$, and
$^1$A$''$ states, respectively; see Figure \ref{fig:n3ppes}.\\

\noindent
For the linear geometry the $^1$A$^{'}$ and $^1$A$^{''}$ PESs are
degenerate. As the middle row of Figure \ref{fig:n3ppes} shows, the
most notable difference between the two singlet states is the presence
of a potential well near $\theta = 90^\circ$ for the $^1$A$^{'}$
state. Thus, a trajectory starting with $\theta \approx 0^\circ$ or
$180^\circ$ on the $^1$A$^{''}$ state remains confined in the
neighbourhood of the potential wells until sufficient energy has
accumulated along the dissociative coordinate(s) to decompose whereas
a trajectory with similar initial conditions but propagating on the
$^1$A$^{'}$ state PES may potentially travel towards $\theta =
90^\circ$ on account of the presence of a minimum at this
position. However, the barrier to reach this minimum is $\sim 2.75$ eV
(see Table \ref{tab:criticalpoints}) which is too high to be overcome
at the conditions studied here. Nevertheless, away from the global
minimum the shapes of the two singlet PESs differ somewhat.\\

\begin{figure}[htbp]
\begin{center}
  \includegraphics[width=1.0\textwidth]{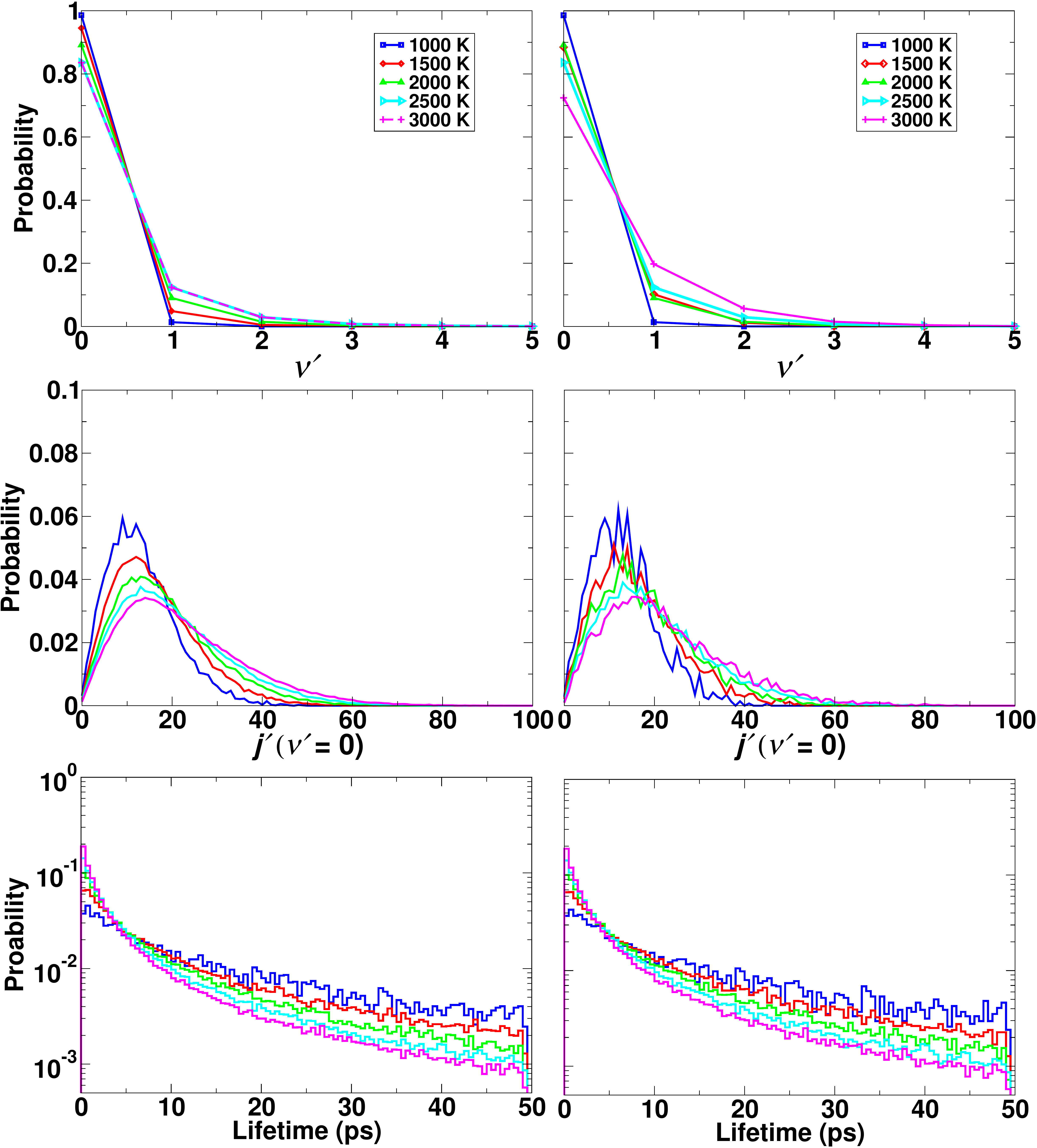}
  	\caption{Photodissociation dynamics for the singlet excited
          state PESs of $\textrm{N}_{3}^{+}$. Product vibrational and
          rotational state distributions and lifetime distributions at
          different temperatures for photodissociation of N$_{3}^{+}$
          on the $^{1}\textrm{A}^{'}$ (left column) and
          $^{1}\textrm{A}^{''}$ (right column) excited states.
          Thermalized initial conditions are generated on the ground
          state at the respective temperatures and propagated on the
          excited state PES. In all cases, the resulting trajectories
          are propagated for a maximum time of 50 ps or until
          dissociation, whichever is earlier.}
\label{fig:final.singlet}
\end{center}
\end{figure}

\noindent
The final state vibrational distributions from dynamics on the
$^1$A$^{'}$ PES are dominated by population of $v'=0$ with maximum
population of $v'=1$ at higher temperatures only reaching $\sim 10$
\%, see left column in Figure \ref{fig:final.singlet}. For the
rotational state distributions $P(j')$ corresponding to $v'=0$ the
maxima occur between $j' = 10$ and $j' = 13$, depending on
temperature. Short lifetimes ($\sim 1$ ps) on the excited state PES
before dissociation are about one order of magnitude more probable
than lifetimes of $\sim 50$ ps at 1000 K. This changes to a difference
of 2 orders of magnitudes at 3000 K with short lifetimes beoming much
more probable.\\

\noindent
For photodissociation from the $^1$A$^{''}$ state (Figure
\ref{fig:final.singlet} right column) excitation of $v'=1$ reaches up
to 20 \% for higher temperatures. This is a clear difference compared
with vibrational products dissociating from the $^1$A$^{'}$ state. For
the rotational distributions corresponding to $v' =0$ the maxima also
shift progressively to higher $j'$ values with increasing temperature
but the maxima occur at somewhat higher rotational quantum numbers
compared with dissociation by populating the $^1$A$^{'}$
state. Finally, for the $^1$A$^{''}$ state at $T \sim 1000$ K, short
lifetimes are about one order of magnitude more probable than long
lifetimes. Short lifetimes become even more probable as temperature
increases. These aspects are similar for dynamics on the $^1$A$^{'}$
state.\\

\noindent
Overall, photodissociation on the two singlet states follows
comparable patterns although details in the final state rotational
distributions indicate that the anisotropy of the $^1$A$^{''}$ state
PES differs somewhat from that of the $^1$A$^{'}$ PES, see middle row
of Figure \ref{fig:n3ppes}. Conversely, photodissociation from the
$^3$A$'$ state leads to more pronounced population of vibrationally
excited states $v' > 0$, in particular at higher temperatures, and the
rotational distributions $P(j'; v'=0)$ appear broader with the maxima
of the distributions shifted to higher values of $j'$ compared with
photodissociation from the two singlet states. This can be related to
the flatter PES along the angular coordinate in the $^3$A$'$ state
compared with the two singlet states which leads to sampling of larger
angular distortions and to increased torque upon photodissociation.\\

\section{Discussion and Conclusion}
So far, the photodissociation dynamics of the N$_3^{+}$ ion in the
lowest singlet and triplet excited states was followed based on QCT
simulations. Consistent with the shape differences between the
$^3\Sigma_{\rm g}^{-}$ ground state PES and the three electronically
excited PESs high rotational excitations in $j'$ after
photodissociation are found. To further corroborate this finding,
simulations starting from sampling the Wigner
distribution\cite{wigner:1932,weinbub:2018} of the $^3 \Sigma_{\rm
  g}^{-}$ ground state wavefunction were also carried out. The Wigner
function $f_{W}(\Gamma)$ related to a three dimensional wavefunction
in Jacobi coordinates $\Psi(R,r,\theta)$ is
\begin{equation}
    f_{W}(\Gamma)=(\pi \hbar)^{-3}\int ds_{R} ds_{r} ds_{\theta} \quad
    e^{2i(Ps_{R}+ps_{r}+Ps_{\theta})/\hbar} \\ \times \Psi
    ^{*}(R+s_{R},r+s_{r},\theta+s_{\theta}) \times
    \Psi(R-s_{R},r-s_{r},\theta-s_{\theta})
    \label{eq:wigner}
\end{equation}
with $\Gamma = (R,r,\theta,P,p,P_\theta)$. Initial conditions for
photodissociation are generated by sampling the probability
distribution $f_{W}(\Gamma)$ in Eq.\ref{eq:wigner} using Metropolis
Monte Carlo importance sampling. Here, $\Psi(R,r,\theta)$ is the
ground state wavefunction for the $^{3}\textrm{A}^{''}$ PES. The
collection of initial conditions generated on the ground state PES in
this manner represent the quantum wavepacket which is then projected
onto the excited state as before and the dynamics is followed from QCT
simulations.\\

\begin{figure}[htbp]
\begin{center}
\includegraphics[width=0.95\textwidth]{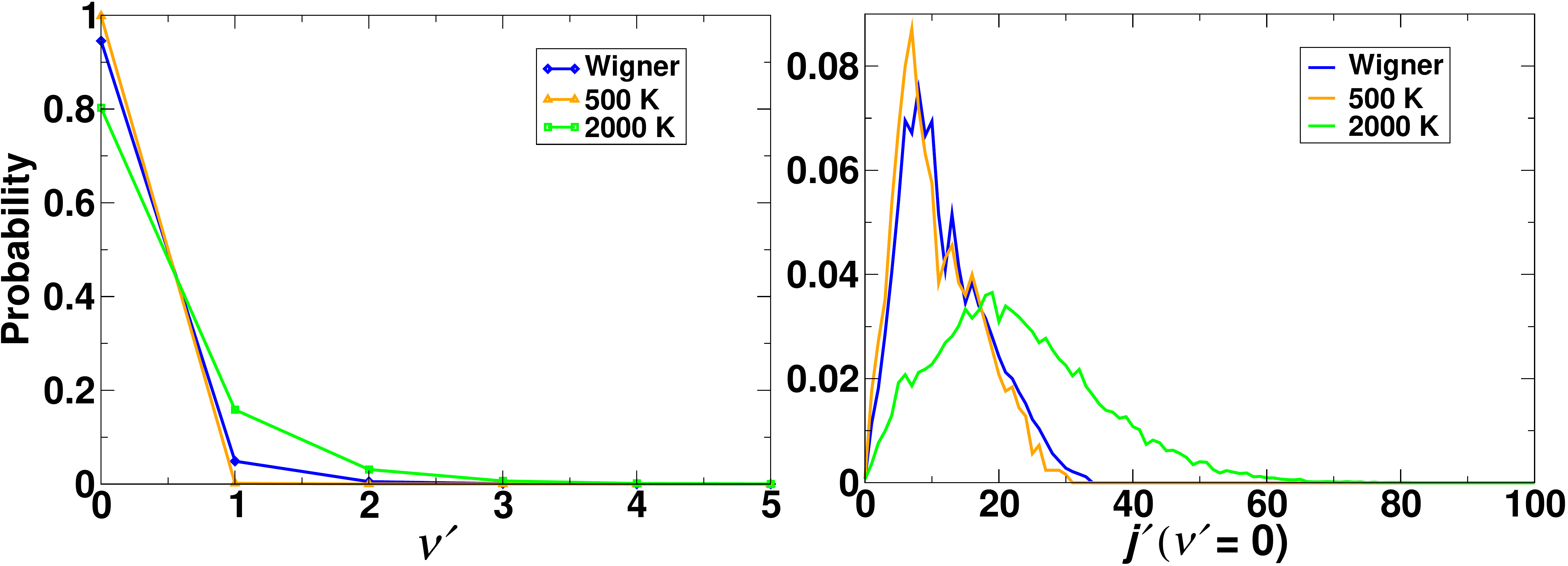}
\caption{Photodissociation dynamics for the $^{3}\textrm{A}^{'}$
  excited state of N$^+_3$ with trajectories started from
  configurations obtained by sampling the Wigner distribution (blue)
  corresponding to the quantum ground state on the $^3 \Sigma_{\rm
    g}^{-}$ state and from QCT simulations at 500 K (orange) and 2000
  K (green), respectively. Final vibrational (left) and rotational
  (right) state distributions are shown. The results are for vertical
  excitation from the ground to the $^3 \Pi_{\rm
    u}$/$^{3}\textrm{A}^{'}$ excited state.}
\label{fig:final_state_wig}
\end{center}
\end{figure}

\noindent
The resulting final state distributions are shown in Figure
\ref{fig:final_state_wig}. The quantum ground state is 1421 cm$^{-1}$
above the minimum energy of the PES, corresponding to 2000 K. Final
state distributions from classical trajectories sampled from the
Wigner distribution follow closely those from classical simulations
with initial conditions at low temperature (500 K). Since
``temperature'' is not a meaningful physical quantity for such small
systems,\cite{schmelzer:2010} $T$ is rather more a label to
distinguish how initial conditions were generated for ensembles which
are increasingly energized. Given this, it is encouraging to see that
final state distributions from classical trajectories starting from
two very different strategies to generate initial conditions are
consistent with one another. \\

\noindent
To corroborate the pronounced coupling of the intermolecular modes it
is also useful to consider the lower quantum bound states for the
different PESs. For this, the 3-dimensional time-independent
Schr\"odinger equation was solved for the lowest bound vibrational
states on the the two lowest singlet electronically excited PESs,
neglecting Renner-Teller coupling. The zero point vibrational energies
for the $^1$A$^{'}$ and $^1$A$^{''}$ PESs are 1809 cm$^{-1}$ and 1707
cm$^{-1}$, respectively. Fundamentals are at 1019 cm$^{-1}$ and 1281
cm$^{-1}$ for the $v_3$ antisymmetric stretch, at 1496 cm$^{-1}$ and
1241 cm$^{-1}$ for the $v_1$ symmetric stretch, and at 1041 cm$^{-1}$
and 814 cm$^{-1}$ for the $v_2$ bending vibration. This compares with
fundamentals at 1096 cm$^{-1}$, 774 cm$^{-1}$, and 395 cm$^{-1}$ for
the $v_1$, $v_3$, and $v_2$ vibrations on the $^3 \Sigma_{\rm g}^{-}$
ground state PES. The higher vibrational states (combination bands and
overtones) are approximately assigned using vibrational quantum
numbers based on node counting; the results are reported in Table
\ref{tab:compstates}. It must be noted that the assignments are rather
approximate due to the anharmonicities and strong couplings prevalent
among the vibrational levels of the excited states. This can also be
seen for the representative wavefunctions for lower energy levels of
the $^1$A$^{'}$ and $^1$A$^{''}$ states shown in Figures
\ref{sifig:n3p_1Ap_wfn} and Figure \ref{sifig:n3p_1App_wfn}
respectively. Similar calculations for the triplet state $^3$A$^{'}$
were not performed as it predissociates and requires full scattering
calculations, which are outside the scope of the present work.\\

\begin{table}[ht!]
 \caption{Lower bound states (in cm$^{-1}$ above the ground state) for
   the singlet electronically excited states of N$^+_3$. The bound
   states are assigned to vibrational quantum numbers $v_1$ (symmetric
   stretch), $v_2$ (bend), and $v_3$ (antisymmetric stretch) by
   node-counting; see Figures \ref{sifig:n3p_1Ap_wfn} and
   \ref{sifig:n3p_1App_wfn} for examples. The labels are only
   approximate due to strong coupling. The angular momentum quantum
   number is $l$.}
\begin{tabular}{l  c |  c c}
  \hline
    $v_1 v_2^l v_3$& $^{1}\textrm{A}'$ & $v_1 v_2^l v_3$& $^{1}\textrm{A}''$\\
  \hline
  \hline
  0 $0^0$ 1   &   1018.8   &  0  $1^1$ 0  &  813.6    \\
0 $1^1$ 0   &   1040.9   &  1  $0^0$ 0  & 1241.2    \\
1 $0^0$ 0   &   1496.1   &  0  $0^0$ 1  & 1281.1    \\
0 $0^0$ 2   &   2033.1   &  0  $2^0$ 0  & 1627.4    \\
0 $1^1$ 1   &   2048.9   &  1  $1^1$ 0  & 2064.1    \\
0 $2^0$ 0   &   2063.3   &  0  $1^1$ 1  & 2068.3    \\
1 $1^1$ 0   &   2490.8   &  0  $0^0$ 2  & 2435.7    \\
1 $0^0$ 1   &   2492.7   &  0  $3^1$ 0  & 2446.8    \\
2 $1^1$ 1   &   2512.8   &  2  $0^0$ 0  & 2455.3    \\
2 $2^0$ 0   &   2825.9   &  1  $0^0$ 1  & 2601.5    \\
2 $0^0$ 0   &   2979.7   &  0  $2^0$ 1  & 2853.4    \\
0 $0^0$ 3   &   3041.1   &  1  $2^0$ 0  & 2885.1    \\
1 $0^0$ 3   &   3481.9   &  2  $1^1$ 0  & 3242.4    \\
\hline
\end{tabular}
\label{tab:compstates}
\end{table}

\noindent
The current work investigates the photodissociation dynamics of the
N$_3^+$ ion for the three lowest electronically excited states. For
this, an ensemble of initial conditions is generated on the ground
state PES and projected onto each of the excited states. For the
$^3$A$^{'}$ PES, it is found that $P(v')$ essentially does not depend
on how the excitation takes place. Excitation of the entire ground
state population gives a similar $P(v')$ compared with near-resonant
excitation or when only the high-energy part of the ground state
distribution is excited. This is somewhat different for $P(j')$ for
which excitation of the high-energy population yields slightly lower
$j_{\rm max}'$ compared with the other three excitation schemes. Using
initial conditions sampled from the Wigner distribution of the ground
state wavefunction to initiate the dynamics on the $^3$A$^{'}$ excited
state leads to comparable final state distributions $P(v')$ and
$P(j')$ as do simulations started from initial conditions generated at
500 K. Product state distributions for the $2 ^3$A$^{''}$ state are
expected to be similar to those from excitation to the $^3$A$^{'}$
state due to the similar shape of the PES for $\theta \leq
45^\circ$. Even for larger bending angles the two PESs are quite
similar except for a high-lying T-shaped minimum for short
$R-$separations which is, however, energetically inaccessible.\\

\noindent
The present work reports testable results for experiments from
classical and semiclassical dynamics on accurate, high-level potential
energy surfaces for this important ion. It is hoped that the
predictions spur experimental efforts to better characterize the
photodissociation dynamics of N$_3^+$. This will be of particular
relevance to atmospheric and interstellar chemistry.\\

\section*{Acknowledgment}
Support by the Swiss National Science Foundation through grants
200021-117810, the NCCR MUST (to MM), and the University of Basel is
also acknowledged. Part of this work was supported by the United
States Department of the Air Force, which is gratefully acknowledged
(to MM). This work was supported by the Australian Research Council
Discovery Project Grants (DP150101427, DP160100474). MM acknowledges
the Department of Chemistry of Melbourne University for a Wilsmore
Fellowship during which this work has been initiated.\\

\section*{Supporting Information}
The supporting information reports the initial conditions on the
ground state PES, individual photodissociating trajectories, the
crossing seams between the singlet PES, and wavefunctions for the
fundamentals on the two singlet PESs.

\section*{Data Availability Statement}
All information necessary to construct the potential energy surfaces
is available at \url{https://github.com/MMunibas/N3p_PESs}.\\

\bibliography{n3p}

\clearpage

\renewcommand{\thetable}{S\arabic{table}}
\renewcommand{\thefigure}{S\arabic{figure}}
\renewcommand{\thesection}{S\arabic{section}}
\renewcommand{\d}{\text{d}}
\setcounter{figure}{0}  
\setcounter{section}{0}  

\noindent
{\bf Supporting Information: Photodissociation Dynamics of
  N$_{3}^{+}$}\\

\begin{figure}[htbp]
\begin{center}
  \includegraphics[width=0.80\textwidth]{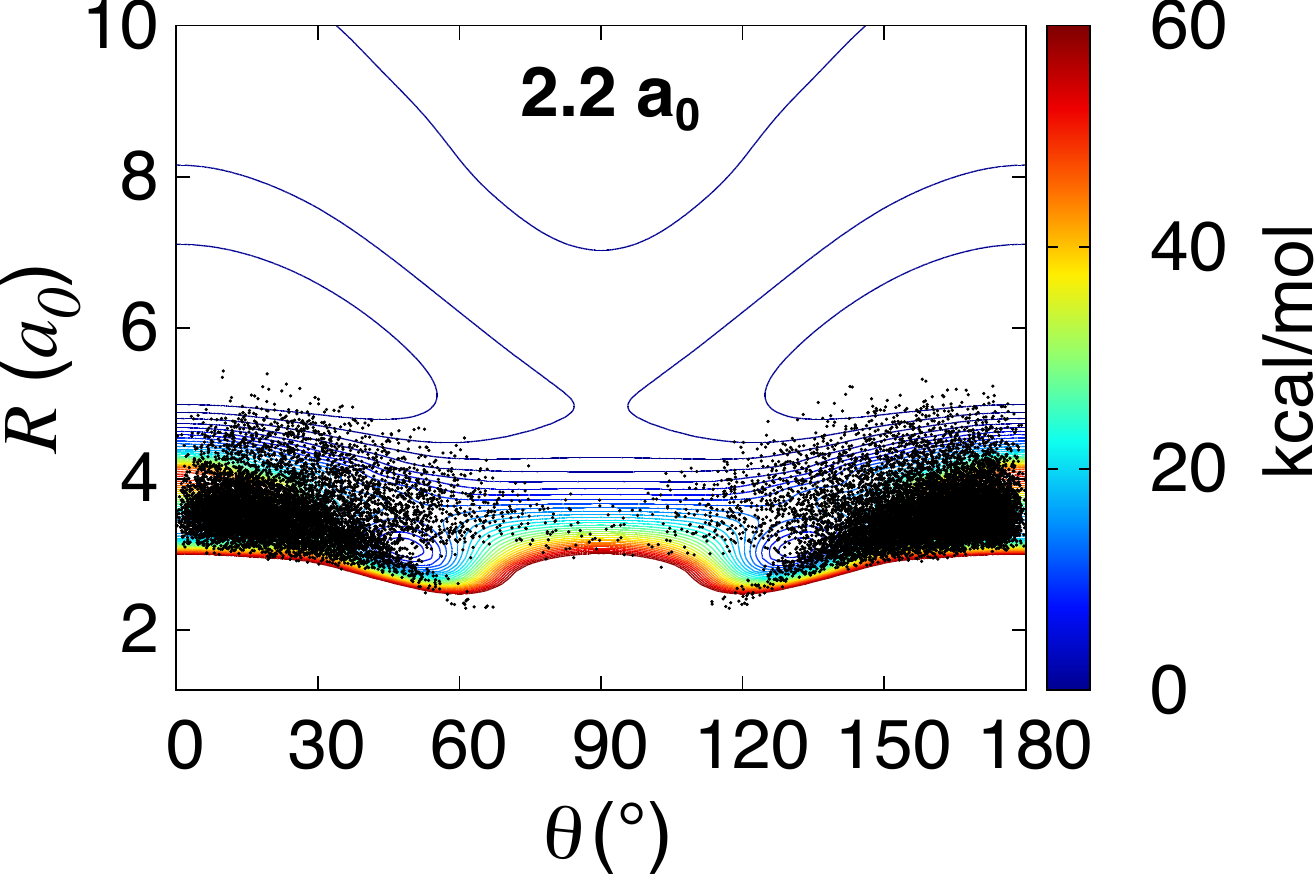}
\caption{500000 configurations generated on the $^{3}\textrm{A}^{''}$
  (ground) state at 1000 K are projected on the $^{3}\textrm{A}^{'}$
  PES of $\textrm{N}_{3}^{+}$. This set of phase space points are used
  as initial conditions for photodissociation. See text for
  discussion.}
\label{sifig:incond_proj_3ap}
\end{center}
\end{figure}

\begin{figure}[htbp]
\begin{center}
\includegraphics[width=0.80\textwidth]{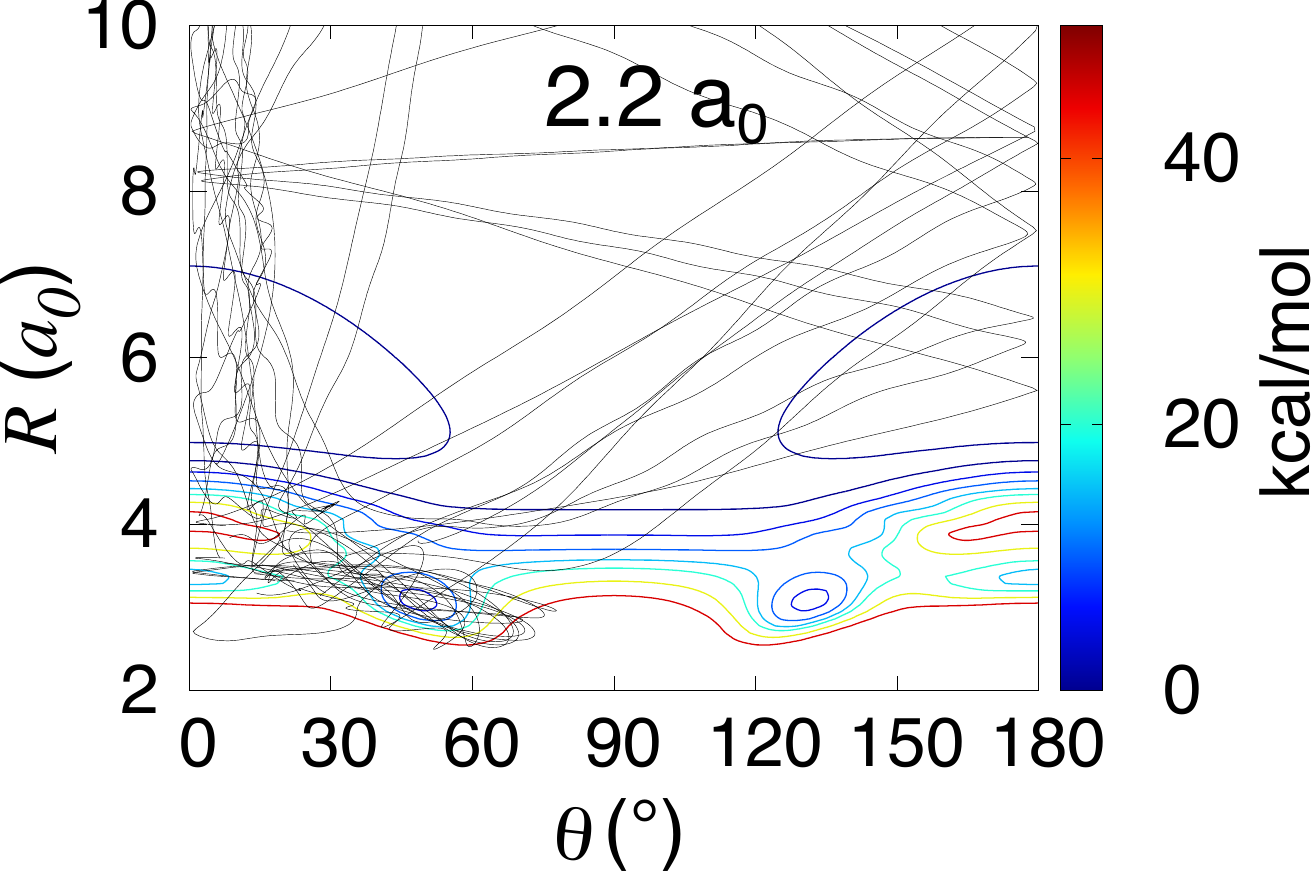}
\caption{Photodissociating trajectories on the $^{3}\textrm{A}^{'}$
  excited state of $\textrm{N}_{3}^{+}$ at 1000 K. Initial conditions
  generated on the $^{3}\textrm{A}^{''}$ (ground) state at 1000 K are
  vertically excited to the $^{3}\textrm{A}^{'}$ state PES and
  propagated up to a final time of 50 ps. The N--N diatomic separation
  is fixed at $r=2.2$ a$_0$ for the underlying PES in the graph.}
\label{sifig:pdtraj_3ap}
\end{center}
\end{figure}

\begin{figure}[htbp]
\begin{center}
\includegraphics[width=1.0\textwidth]{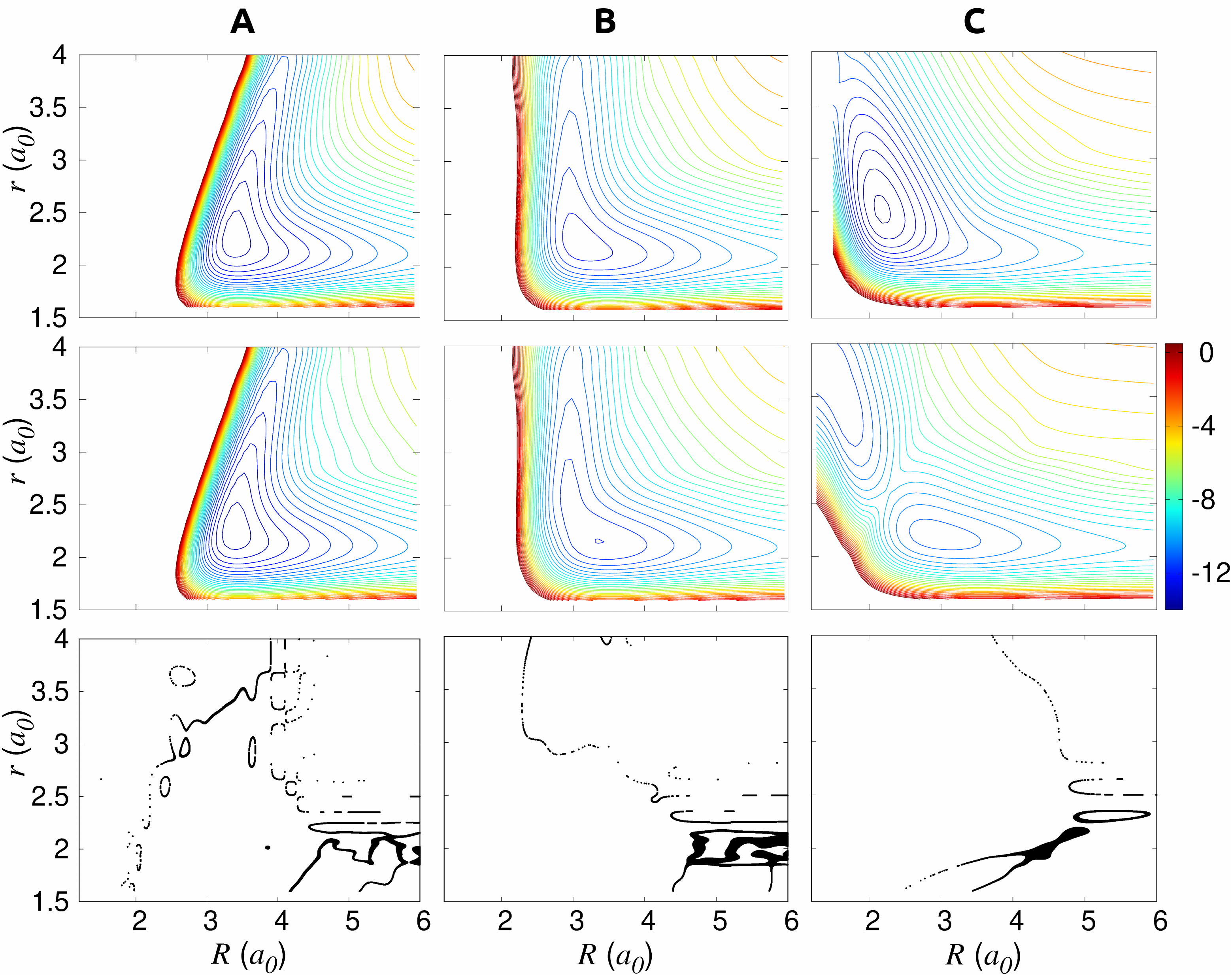}
\caption{Contour plots $V(R,r)$ for the singlet excited states of
  N$_3^+$ and the crossing points between them. Configurations are
  considered on the crossing seam if the energy difference between the
  two PESs is within $10^{-5}$ eV. The surfaces are shown for three
  different values of the Jacobi angle $\theta=0.01^\circ$ (panel A),
  $\theta=45^\circ$ (panel B), and $\theta=90^\circ$ (panel C). The
  top and middle rows show the $^1$A$^{'}$ and $^1$A$^{''}$ PESs
  respectively while the bottom row shows the points at which the
  respective PESs cross each other.  All energies are in eV.}
	\label{sifig:n3_crossing}
\end{center}
\end{figure}

\begin{figure}[htbp]
\begin{center}
\includegraphics[width=0.90\textwidth]{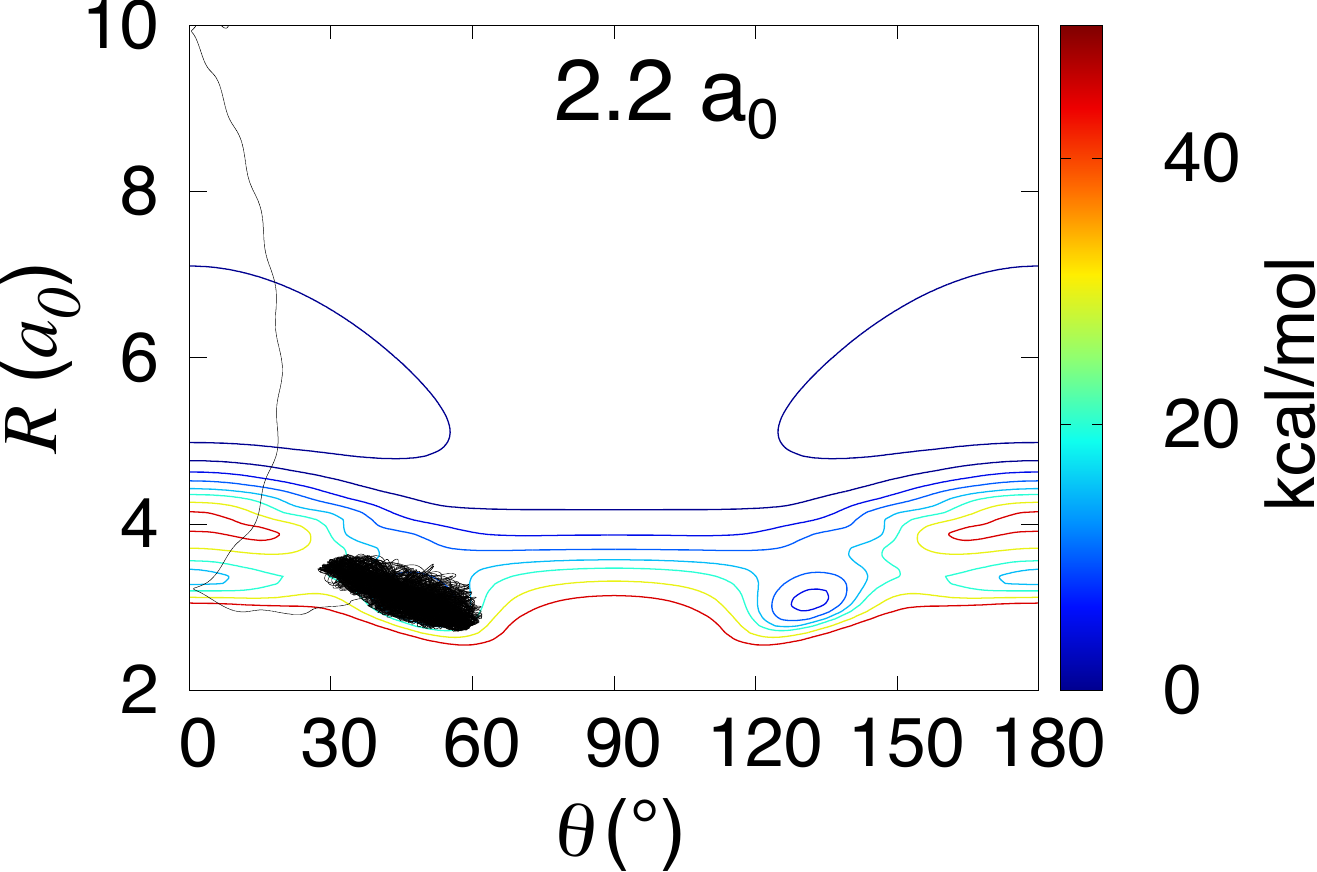}
\caption{A trajectory with an excited state lifetime of $\sim17$ ps
  projected on the $^{3}\textrm{A}^{'}$ PES. The simulations were
  carried out at 1000 K. The 2d-PES is shown for $r=2.2$ a$_0$. }
\label{sifig:long_lifetime_3Ap}
\end{center}
\end{figure}

\begin{figure}[htbp]
\begin{center}
\includegraphics[width=1\textwidth]{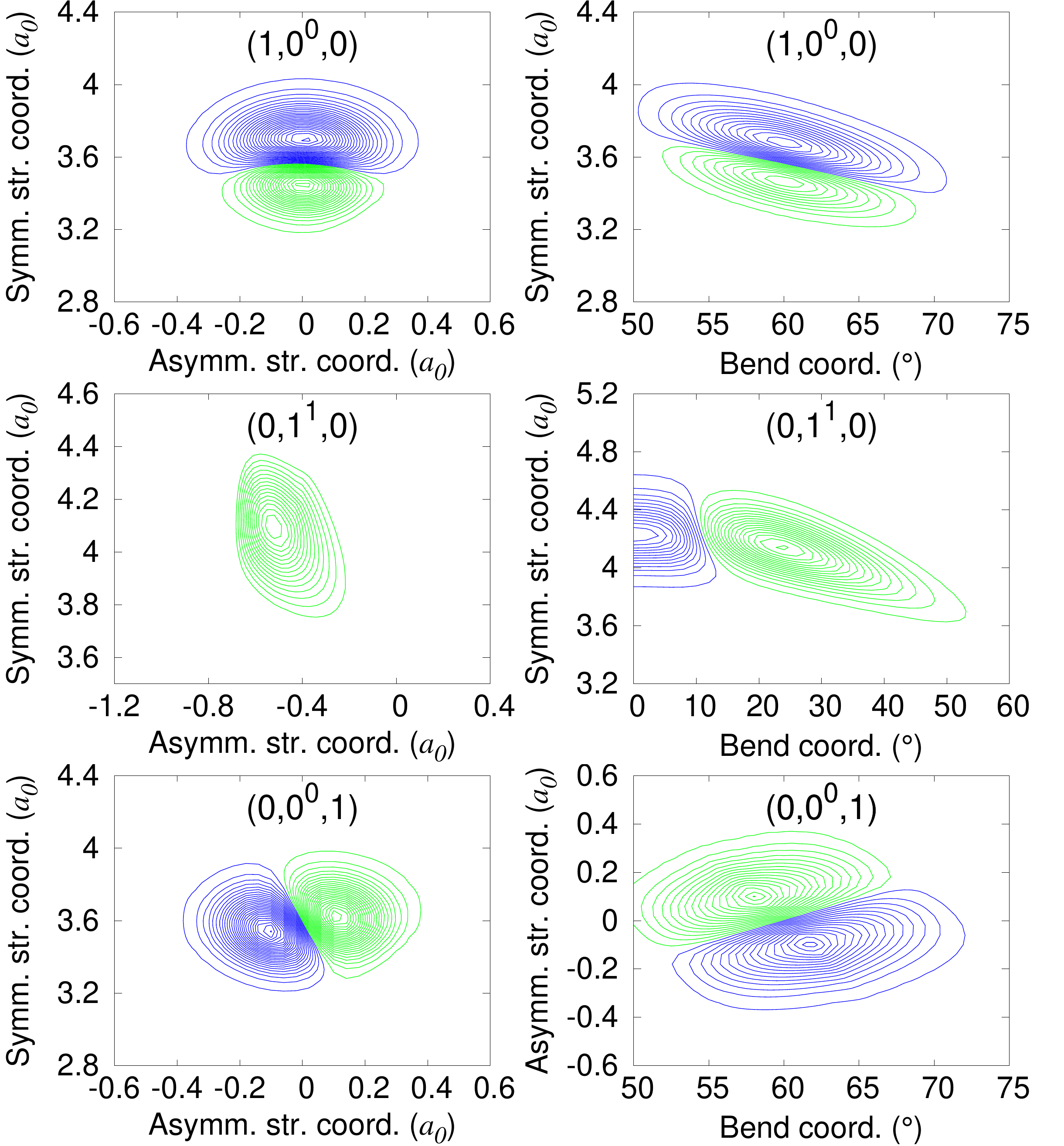}
\caption{Wave functions for the fundamentals of $^{1}\textrm{A}^{'}$
  excited state of $\textrm{N}_{3}^{+}$ from DVR3D calculations. The
  states are indicated by the labels ($\nu_1 \nu_2^l \nu_3$).}
\label{sifig:n3p_1Ap_wfn}
\end{center}
\end{figure}

\begin{figure}[htbp]
\begin{center}
\includegraphics[width=1\textwidth]{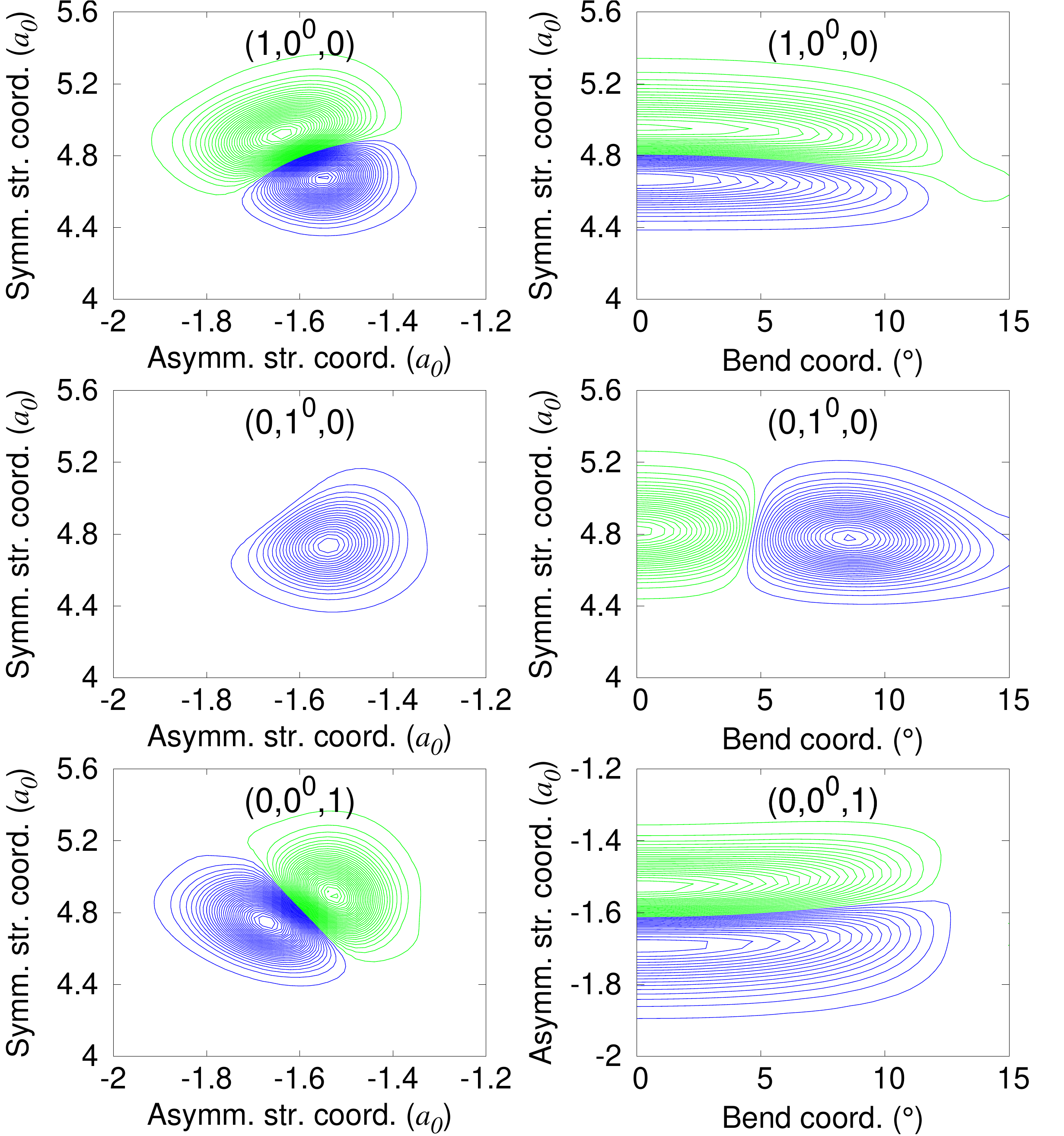}
\caption{Wave functions for the fundamentals of $^{1}\textrm{A}^{''}$
  excited state of $\textrm{N}_{3}^{+}$ from DVR3D calculations. The
  states are indicated by the labels ($\nu_1 \nu_2^l \nu_3$).}
\label{sifig:n3p_1App_wfn}
\end{center}
\end{figure}

\end{document}